**Tables now available at the following address :**

**http://www.lcpmr.upmc.fr/themes-A2f.php**

# Cauchois & Sénémaud Tables of wavelengths of x-ray emission lines and absorption edges

Philippe Jonnard and Christiane Bonnelle

*Laboratoire Chimie Physique – Matière Rayonnement, UPMC Univ Paris 06, CNRS UMR 7614, 11 rue Pierre et Marie Curie, F-75231 Paris cedex 05, France*

We present the Cauchois & Sénémaud Tables of x-ray emission lines and absorption edges. They are written both in French and English. They were published in 1978 by Pergamon Press and are insufficiently known. However they are of large interest because of their completeness. They comprise the energies of all the K, L, M, N and O emission lines of natural elements from lithium up to uranium as well as the energies of satellite emissions and absorption discontinuities. The more intense lines of radio-elements up to fermium (Z=100) are also given. The Tables range from the hard x-rays (122 keV, 0.01 nm) to the extreme ultra-violet (30 eV, 42 nm). For each transition, the wavelength (Å and uX) and energy (eV and Ry) are given and references are indicated. The transitions are grouped by increasing wavelength (decreasing photon energy) and also by element and spectral series. We present as an example the use of the Tables to identify the emissions of the molybdenum L spectrum. We decided to scan the Cauchois & Sénémaud Tables and make them available for the scientific community. They are now available at the Website of our laboratory, http://www.lcpmr.upmc.fr/.

**Corresponding author**: Dr. Philippe Jonnard, Laboratoire Chimie Physique – Matière Rayonnement, 11 rue Pierre et Marie Curie, F-75231 Paris cedex 05, France

Philippe.jonnard@upmc.fr tel: 33 1 44 27 66 08 fax: 33 1 44 27 62 26

## 1. Introduction

The tables of wavelengths of x-ray emission lines and absorption edges (C&S Tables hereafter) [1] presented in this paper were established by Pr. Yvette Cauchois and Dr. Christiane Sénémaud, Figure 1, and published as a book in 1978 by Pergamon Press (now Elsevier). Both Pr. Cauchois and Dr. Sénémaud have made their whole scientific career at the Laboratoire de Chimie Physique of the Pierre et Marie Curie University and CNRS in Paris. Pr. Cauchois [2] was famous worldwide for her outstanding contributions in x-ray physics and x-ray spectroscopy, particularly for the high resolution bent crystal spectrometer working in transmission (Cauchois spectrometer) [3] and for its pioneer work to use the synchrotron radiation as a bright x-ray source [4,5]. Dr. Sénémaud [6], used x-ray spectroscopy techniques, and particularly x-ray emission spectroscopy, to study the electronic structure of solids. She also took part in the development of x-ray spectrometers [7,8].

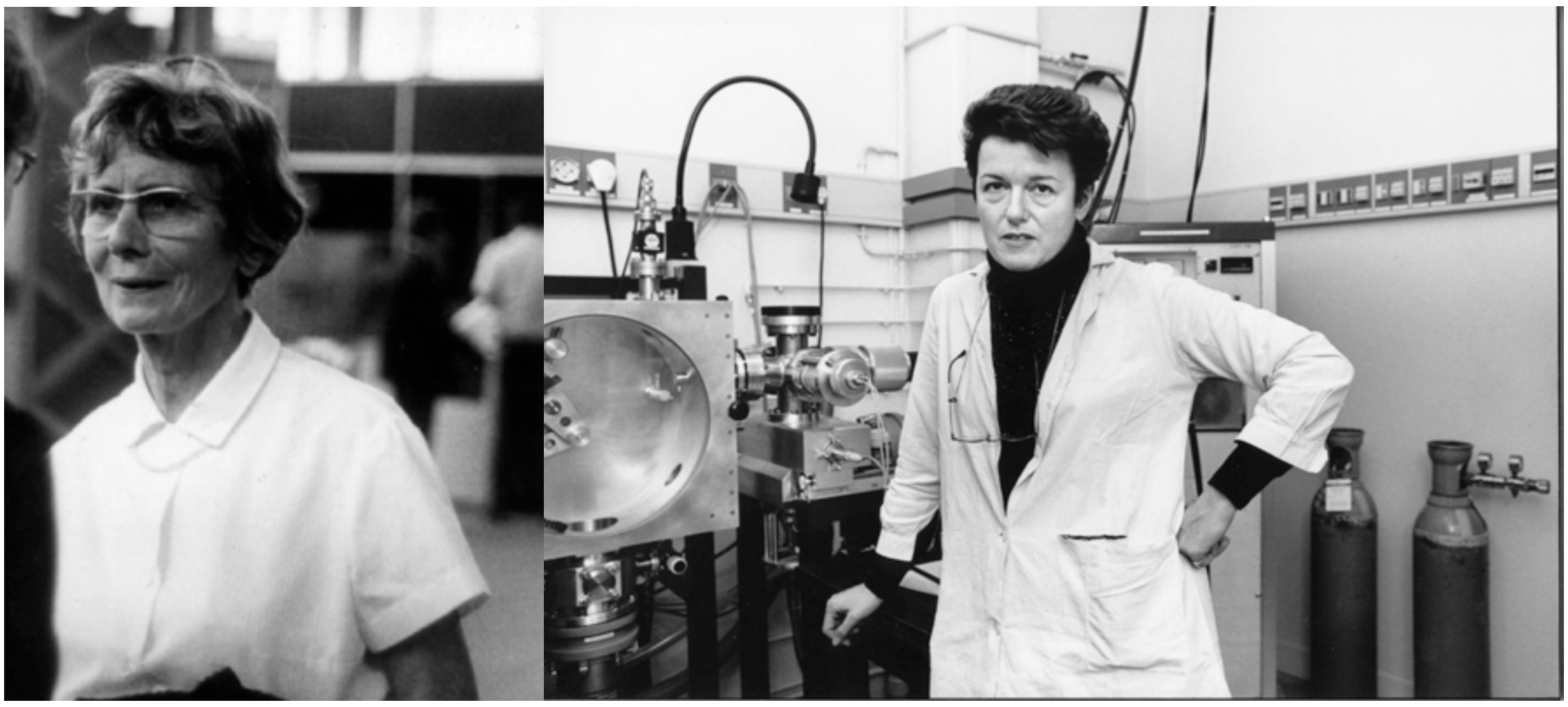

*Figure 1: Professor Yvette Cauchois (left) and doctor Christiane Sénémaud (right).*

One of the motivations of establishing of the C&S Tables was to update and complete previous tables published in 1947 by Y. Cauchois and H. Hulubei [9]. The interest of the C&S Tables comes from their completeness since they give the wavelengths of the emission lines, main and satellites, and of the absorption edges, for all the elements and also for some radio-elements up to Z=100. Thus, this is an all-in-one book where users of x-ray spectrometers can find useful information. In contrast, the x-ray data booklet [10], widespread in the community of the synchrotron radiation users, presents only the most intense lines of the K, L and M series. These values are in fact extracted from the Bearden and Burr tables [11]. These tables are rather complete but only give the wavelengths of the diagram emission lines. Another complete set of x-ray transitions can be found in a recent handbook [12]. One more recent publication of tables exists [13] and is available from the NIST website [14]. It gives wavelengths for elements up to Z=100 for K and L diagram lines with precise values. Indeed, the experimental values are on a scale consistent with the International System of measurement and the numerical values are determined using constants from the Recommended Values of the Fundamental Physical Constants. Moreover, accurate theoretical estimates are included for all transitions.

## 2. Presentation of the Cauchois & Sénémaud Tables

The C&S Tables were written both in French and English. The presented data were collected from experimental values published until 1977. We show in Figure 2 the English table of contents.

Wavelengths of
X-RAY EMISSION LINES
and
ABSORPTION EDGES

*C O N T E N T S*

*Figure 2: The table of contents of the C&S Tables.*

The first part of the C&S Tables, pages 1i-3i, is an introduction that presents the way the experimental data were collected from the literature. It also gives a brief description of the Tables, the used notations and abbreviations and some conversion factors.

The second part, pages 1-255, gives the wavelengths by increasing order, from the hard x-rays (0.01 nm, 122 keV) to the extreme UV (102 nm, 12.1 eV) for the main or diagram lines, i.e. the lines that can be described by a transition in a singly ionized system. For some rare earths, transitions in the mono-excited systems are also given. The values are presented for the pure elements except in some cases where emissions from compounds are chosen and indicated in a note. The K, L, M, N and O lines of all the elements can be found. The absorption edges are also indicated, up to plutonium (Pu, Z=94) for the K edges, americium (Am, Z=95) for the L edges, plutonium for the M edges, holmium (Ho, Z=67) for the N edges and bismuth (Bi, Z=83) for the O edges. An example presenting page 197 is shown in Figure 3. In this part the given informations are :

- the wavelength in uX ("unité X" or "X unit", 1 mÅ ≈ 1.002 uX) or kuX depending on the photon wavelength range;
- the wavelength in Å ($10^{-10}$ m) or mÅ ($10^{-13}$ m);
- the element and its atomic number;
- the electron transition in IUPAC notation [15], i.e. mentioning the initial and final levels of the transition, or the level for an absorption edge;
- for the most intense lines, their name in Siegbahn notation [15];
- the diffraction order at which the line appears;
- the photon energy in eV or keV;
- the wave-number divided by the Rydberg constant, which is equivalent to the energy in Ry (1 Ry = 13.6 eV);
- the square root of the wave-number divided by the Rydberg constant, used to plot Moseley curves;
- the bibliographic reference for the emission line.

Raies de diagramme: 6058,0–6508

Notations, p. 31 — Notes, p. 254. Références bibliographiques, p. 299

| λ/uX | λ/mÅ | Elément | Transition | Notation usuelle | Ordre | E/keV | ν/R | $(\nu/R)^{1/2}$ | Bibl. |
|---|---|---|---|---|---|---|---|---|---|
| 6058,0 | 6070,7 | 40 Zr | $L_{III}M_V$ | $L\alpha_1$ | I | 2,0423 | 150,11 | 12,2519 | (41.1) |
| 6065,3 | 6078,0 | 40 Zr | $L_{III}M_{IV}$ | $L\alpha_2$ | I | 2,0399 | 149,93 | 12,2445 | (41.1) |
| 6076 | 6089 | 74 W | $M_{III}N_V$ | $M\gamma$ | I | 2,036 | 149,7 | 12,234 | (31.3) |
| 6081,7 | 6094,5 | 39 Y | $L_{III}N_I$ | $L\beta_6$ | I | 2,0344 | 149,52 | 12,2280 | (41.1) |
| 6121 | 6134 | 74 W | $M_{III}N_{IV}$ | – | I | 2,021 | 148,6 | 12,189 | (31.3) |
| 6138,1 | 6151,0 | 42 Mo | $L_{III}M_I$ | $L\ell$ | I | 2,0157 | 148,15 | 12,1717 | (41.1) |
| 6144,0 (h) | 6156,9 | 15 P | $KL_{III}$ | $K\alpha_1$ | I | 2,0138 | 148,01 | 12,1658 | (63.21) |
| 6146,7 (h) | 6159,6 | 15 P | $KL_{II}$ | $K\alpha_2$ | I | 2,0129 | 147,94 | 12,1631 | (63.21) |
| 6149 | 6162 | 83 Bi | $M_{IV}N_{III}$ | – | I | 2,012 | 147,9 | 12,161 | (31.3) |
| 6159,9 | 6172,8 | 38 Sr | abs. $L_{II}$ | – | I | 2,0086 | 147,63 | 12,1501 | (35.10) |
| 6194 | 6207 | 76 Os | abs. $M_V$ | – | I | 1,998 | 146,8 | 12,117 | (8.277) |
| 6198,1 | 6211,1 | 41 Nb | $L_{II}M_I$ | $L\eta$ | I | 1,9962 | 146,72 | 12,1126 | (41.1) |

*Figure 3: Extract from page 197 of the C&S Tables, where wavelengths are classified in increasing order.*

For the trans-uranium elements, the number of lines is limited. However, the K lines are given for elements up to fermium (Fm, Z=100), the L lines for elements up to curium (Cm, Z=96) and the M lines for elements up to plutonium. Regarding the satellite lines, the covered photon energy ranges from 0.42 nm (30 keV) to 42 nm (29 eV) and spans elements from lithium (Li, Z=3) to antimony (Sb, Z=51) for the K lines, from magnesium (Mg, Z=12) to uranium (U, Z=92) for the L lines and from scandium (Sc, Z=21) to uranium for the M lines.

The third part of the C&S Tables, pages 256-297, gives the emission and absorption wavelengths for the elements of the periodic table in increasing atomic number. The page 260, Figure 4 , is shown as an example giving the K emissions of the elements between lithium and titanium (Ti, Z=22). When the emission is a band (transition from valence states to core levels), this is notified. All the wavelengths in this part can be also found in the previous part. The given informations are :

- the wavelength in Å, mÅ, uX or kuX;
- the element and its atomic number;
- the electron transition in IUPAC notation;
- for the most intense lines, their name in Siegbahn notation.

EMISSIONS DE DIAGRAMME – DIAGRAM EMISSION LINES AND BANDS

SÉRIE K

- 260 -

Série K - 2 He +

| Élément | K | | | | | | | | | | | | | | Élément |
|---|---|---|---|---|---|---|---|---|---|---|---|---|---|---|---|
| | $L_I$ | $L_{II}$ | $L_{III}$ | $M_{II}$ | $M_{III}$ | $M_{IV}$ | $M_V$ | $N_{II}$ | $N_{III}$ | $N_{IV}$ | $N_V$ | $O_{II}$ | $O_{III}$ | O,P | |
| | | $\alpha_2$ | $\alpha_1$ | $\beta_3$ | $\beta_1$ | $\beta_5^{II}$ | $\beta_5^{I}$ | $\beta_2^{II}$ | $\beta_2^{I}$ | $\beta_4^{II},\beta_{4x}$ | $\beta_4^{I}$ | | | | |
| 2 He | | Bande Kα (Maxima) | | | | | | | | | | | | | 2 He |
| 3 Li | | 228,6 | | | | | | | | | | | | | 3 Li |
| 4 Be | | 113,53 | | | | | | | | | | | | | 4 Be |
| 5 B | | 66,95 | | | | | | | | | | | | | 5 B |
| 6 C | | 44,79 | | | | | | | | | | | | | 6 C |
| 7 N | | 31,57 | | | | | | | | | | | | | 7 N |
| 8 O | | 23,55 | | Bande Kβ (Maxima) | | | | | | | | | | | 8 O |
| 9 F | | 18,312 | | | | | | | | | | | | | 9 F |
| 10 Ne | | 14,6154 | | 14,460 | | | | | | | | | | | 10 Ne |
| 11 Na | | 11,9095 | | 11,618 | | | | | | | | | | λ/Å | 11 Na |
| 12 Mg | | 9869,9 | | 9539 | | | | | | | | | | λ/uX | 12 Mg |
| 13 Al | | 8324,10 | 8322,00 | 7965 | | | | | | | | | | | 13 Al |
| 14 Si | | 7113,39 | 7110,66 | 6738,5 | | | | | | | | | | | 14 Si |
| 15 P | | 6146,7 | 6144,0 | | 5792,1 | | | | | | | | | | 15 P |
| 16 S | | 5363,66 | 5360,83 | 5019,4 | | | | | | | | | | | 16 S |
| 17 Cl | | 4720,96 | 4718,07 | 4395,01 | | | | | | | | | | | 17 Cl |
| 18 Ar | | 4186,10 | 4183,17 | | | | | | | | | | | | 18 Ar |
| 19 K | | 3737,07 | 3733,68 | | 3446,80 | 3434,5 | | | | | | | | | 19 K |
| 20 Ca | | 3354,81 | 3351,48 | 3083,2 | | 3068,2 | | | | | | | | | 20 Ca |
| 21 Sc | | 3028,40 | 3025,03 | | 2773,9 | 2758,0 | | | | | | | | | 21 Sc |
| 22 Ti | | 2746,505 | 2742,869 | 2508,744 | | 2493,0 | | | | | | | | | 22 Ti |

*Figure 4: Extract from page 260 of the C&S Tables, where wavelengths are given for each element.*

The last part, pages 298-320, is the bibliography and an index of the authors. It contains 770 references ranging from 1926 to 1977 and about 600 authors. The last page is an addendum made by the authors of this paper that corrects some typing errors and adds some missing lines.

## 3. Example of identification of a x-ray spectrum from the C&S Tables

We present as an example the identification of the lines of the Mo L spectrum from the C&S Tables. A bulk molybdenum (Mo, Z=42) sample was used as a target. The Mo atoms were excited by electron bombardment. The electron energy was set to 6 keV. The radiation emitted from the sample was dispersed by a InSb (111) crystal (reticular distance equal to 0.374 nm) of a high-resolution bent-crystal x-ray spectrometer (Johann-type).

The Mo L spectrum is presented in Figure 5. It ranges from about 2000 eV to 2850 eV. No correction to the efficiency of the detection and to the crystal reflectivity as a function of the photon energy has been applied, because we are interested here only in the energies of the emission lines and not in their intensities. This spectrum has a shape close to the one of zirconium (Zr, Z=40) observed in comparable experimental conditions [16]: with a high-resolution spectrometer and upon electron irradiation. Twenty structures are observed, divided into three classes and given in Table 1:

- Main, or diagram, lines corresponding to one-electron transitions, all identified in the C&S Tables;
- satellites lines, corresponding to transitions in multiply ionised atoms, observed as an asymmetry toward the high photon energy side of the main lines, all identified in the C&S Tables, except those of the Ll and Lη lines; but these lines are very weak (about 100 times less intense than the Lα line) and this explains that their satellites have not been reported yet;
- unidentified lines, quoted in (3) and (6); these very weak lines may be high energy satellites of the Ll and Lη lines, because the energy relative to that of the main line is almost the same in both cases; the same satellite is not observed for the Lα and $L\beta_1$ lines because of the higher background in this case between the Lα and $L\beta_1$ lines and because the presence of the $L\beta_{4,6}$ line above the $L\beta_1$ line; these unidentified lines could also be ascribed to radiative Auger satellites of the Lα and $L\beta_1$ lines; it is not surprising that these faint lines have not been previously reported.

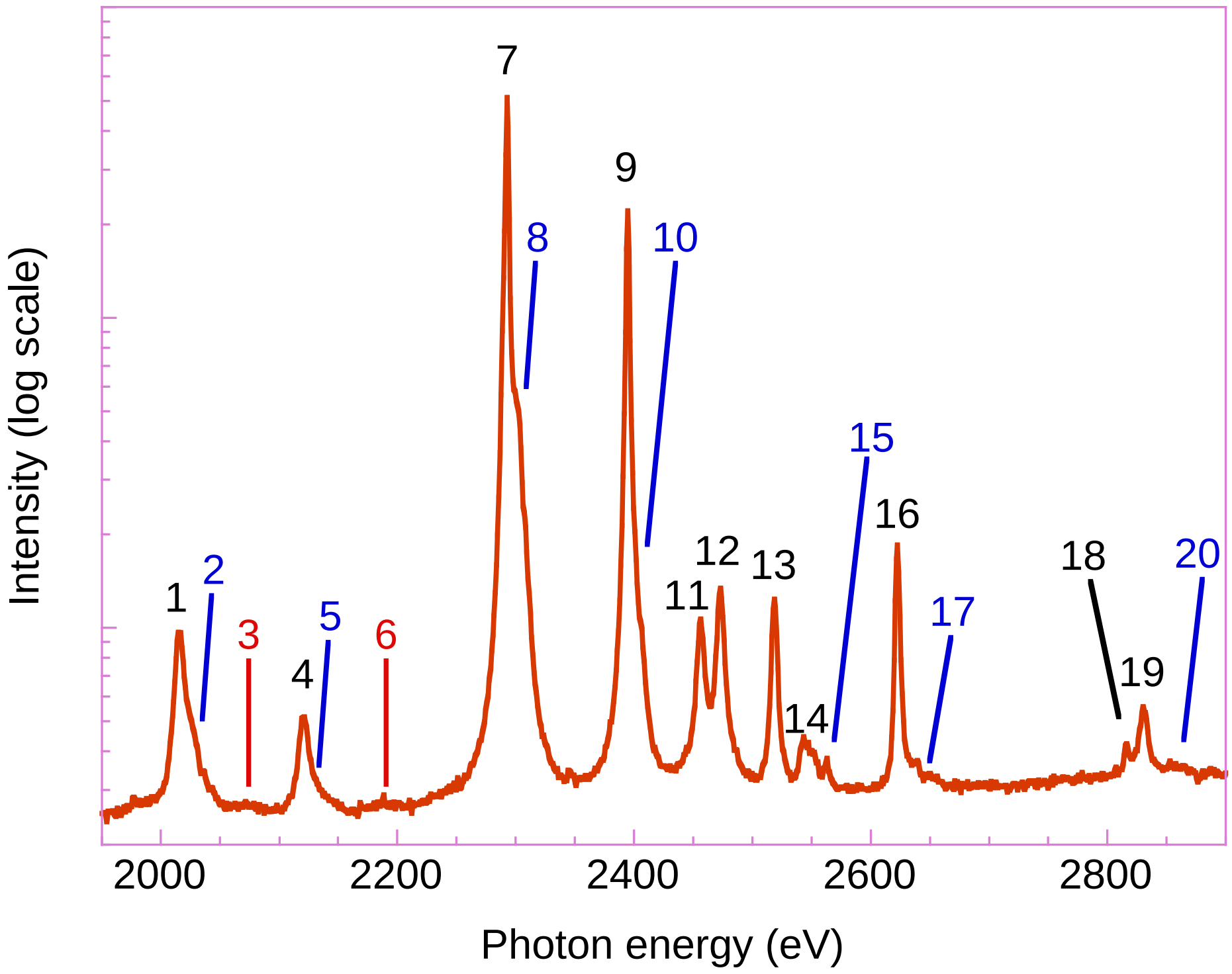

*Figure 5: Mo L spectrum of Mo metal obtained with a high-resolution x-ray spectrometer. The lines are numbered from 1 to 20 (see Table I): in black, main lines; in blue, satellite lines; in red, unidentified lines.*

*Table 1: Identification of the lines observed in the Mo L spectrum (see Figure 5) and comparison of the experimental and C&S Tables values of the emission energies E. For the satellite emissions the energy difference ΔE with respect to the main line is indicated.*

| Number | Name | Transition | E (eV) / ΔE(eV) | E C&S (eV) |
|---|---|---|---|---|
| 1 | Ll | 3s - $2p_{3/2}$ | 2015.2 | 2015.7 |
| 2 | Sat Ll | | 9.1 | Not in Tables |
| 3 | Sat Ll ? | | 61 | Not in Tables |
| 4 | Lη | 3s - $2p_{1/2}$ | 2120.5 | 2119.7 |
| 5 | Sat Lη | | 10.5 | Not in Tables |
| 6 | Sat Lη ? | | 64 | Not in Tables |
| 7 | $L\alpha_{1,2}$ | $3d_{5/2}$ - $2p_{3/2}$<br>$3d_{3/2}$ - $2p_{3/2}$ | 2292.8<br>2289.6 | 2293.19<br>2289.88 |
| 8 | Sat Lα | | 9.3 | 4 - 30 |
| 9 | $L\beta_1$ | $3d_{3/2}$ - $2p_{1/2}$ | 2394.7 | 2394.83 |
| 10 | Sat $L\beta_1$ | | 10.7 | 5 - 10 |
| 11 | $L\beta_{4,6}$ | $3p_{1/2}$ – 2s<br>4s – $2p_{3/2}$ | 2456.2 | 2458.4 |
| 12 | $L\beta_3$ | $3p_{3/2}$ – 2s | 2473.0 | 2473.0 |
| 13 | $L\beta_2$ | 4d – $2p_{3/2}$ | 2518.7 | 2518.3 |
| 14 | Sat $L\beta_2$ | | 25 - 35 | 25 - 33 |
| 15 | $L\gamma_5$ | 4s – $2p_{1/2}$ | 2563.3 | 2563.2 |
| 16 | $L\gamma_1$<br>Cl $K\alpha_{1,2}$ | 4d – $2p_{1/2}$<br>$2p_{3/2,1/2}$ – 1s | 2622.5 | 2623.5 |
| 17 | Sat $L\gamma_1$ | | 29 | 30 |
| 18 | Cl Kβ | 3p – 1s | 2816.4 | 2815.1 |
| 19 | $L\gamma_{2,3}$ | $4p_{3/2,1/2}$ – 2s | 2830.7 | 2830.7 |
| 20 | Sat $L\gamma_{2,3}$ | | 20 | 30 - 35 |

The comparison of the Mo L spectrum to the emissions from the C&S Tables gives evidence of a supplementary line (18). In fact, this emission is ascribed to the Cl Kβ emission from a chlorine impurity present at the surface or in the bulk of the sample. The Cl Kα line is merged with the Mo $L\gamma_1$ emission (16). This should explain why this γ emission is more intense than expected.

The comparison of the experimental and tabulated emission is quite good for the main and satellite lines. The difference is generally a few tenths of eV. Differences larger than 0.5 eV are noted for weak lines (5) and (17), or lines including double contributions (11) and (16). The only exception is line (20), the satellite of the $L\gamma_{2,3}$ emission, where a discrepancy of 10 eV is noted.

It is shown on this example the interest to have tables compiling the whole main and satellite lines.